\documentclass{jfm}

\pdfoutput=1

\usepackage{graphicx}
\usepackage{xcolor}
\usepackage{natbib}
\usepackage{amsmath}
\usepackage{bm}

\newcommand{\ie}{i.e.,\ }
\newcommand{\eg}{e.g.,\ }

\newcommand{\figrefC}[2][]{\mbox{Figure \ref{#2}(#1)}}
\newcommand{\figrefAC}[1]{\mbox{Figure \ref{#1}}}

\newcommand{\equref}[1]{\mbox{equation (\ref{#1})}}
\newcommand{\figref}[2][]{\mbox{figure \ref{#2}(#1)}}
\newcommand{\figrefA}[1]{\mbox{figure \ref{#1}}}
\newcommand{\tabref}[1]{\mbox{table \ref{#1}}}
\newcommand{\tabrefc}[1]{\mbox{Table \ref{#1}}}

\definecolor{cinnamon}{rgb}{0.82, 0.41, 0.12}

\newcommand{\aver}[1]{\left\langle {#1} \right\rangle}

\title[Turbulent channel flow over an anisotropic porous wall]
{Turbulent channel flow over an anisotropic porous wall - Drag increase and reduction}

\author[M.E. Rosti, L. Brandt and A. Pinelli]{Marco E. Rosti$^1$\thanks{Email address for correspondence: merosti@mech.kth.se}, Luca Brandt$^1$ and Alfredo Pinelli$^2$}

\affiliation{$^1$ Linn\'{e} Flow Centre and SeRC, KTH Mechanics, Stockholm, Sweden\\$^2$ School of Mathematics, Computer Science and Engineering. City, University of London, UK}

\begin{document}

\maketitle

\begin{abstract}

The effect of the variations of the permeability tensor on the close-to-the-wall behaviour of a turbulent channel flow bounded by porous walls is explored using a set of direct numerical simulations. It is found that the total drag can be either reduced or increased by more than $20\%$ by adjusting the permeability directional properties. Drag reduction is achieved for the case of materials with permeability in the vertical direction lower than the one in the wall-parallel planes. This configuration limits the wall normal velocity at the interface while promoting an increase of the tangential slip velocity leading to an almost ``one-component" turbulence where the low- and high-speed streaks coherence is strongly enhanced. On the other hand, strong drag increase is found when a high wall-normal and low wall-parallel permeabilities are prescribed. In this condition, the enhancement of the wall-normal fluctuations due to the reduced wall-blocking effect triggers the onset of structures which are strongly correlated in the spanwise direction, a phenomenon observed by other authors in flows over isotropic porous layers or over ribletted walls with large protrusion heights. The use of anisotropic porous walls for drag reduction is particularly attractive since equal gains can be achieved at different Reynolds numbers by rescaling the magnitude of the permeability only.
\end{abstract}

\section{Introduction} \label{sec:introduction}
One of the central problems in fluid engineering concerns the fact that wall-bounded turbulent flows usually exert a much higher wall friction than laminar ones. For this reason, many researchers have studied the flow over surfaces of different characteristics and in particular over porous walls as it is commonly found in many industrial and natural flows. The main aim of these studies was to investigate the effects of locally modified wall boundary conditions on the flow field and to inspire the design of novel surfaces able to deliver technological benefits by altering the near-to-the-wall flow behaviour. In this context, this work explores the interactions between a turbulent shear flow and a permeable porous wall and demonstrates that it is possible to engineer a porous medium of anisotropic permeability to obtain  favourable conditions such as drag reduction or enhanced mixing.

Research in the past has mainly focused on porous surfaces with isotropic permeability that have as a main effect the destabilization of the mean flow and the enhancement of the Reynolds shear stresses with a consequent increase in skin-friction drag in wall bounded turbulent flows. \citet{beavers_sparrow_magnuson_1970a} provided the first experimental evidence that an isotropic porous wall played a destabilising role on the mean velocity field. More recently, \citet{tilton_cortelezzi_2006a, tilton_cortelezzi_2008a} have performed a three-dimensional temporal linear stability analysis of a laminar channel flow bounded by an isotropic porous wall proving that wall permeability can drastically decrease the stability of the flow. \citet{breugem_boersma_uittenbogaard_2006a} have numerically studied the influence of a highly permeable porous wall, made of a packed bed of spheres, on a turbulent channel flow. The results show that the structure and the dynamics of turbulence are quite different from those of a canonical turbulent flow over a smooth, impermeable wall. In particular, it turns out that low- and high-speed streaks and the coexisting quasi-streamwise vortices loose intensity as compared to the solid wall case. Moreover, in agreement with other authors, they report the presence of large spanwise vorticity structures that contribute to increase the exchange of momentum between the porous medium and the channel, thus inducing a strong increase in the Reynolds-shear stresses and, consequently, in the resulting skin friction. \citet{rosti_cortelezzi_quadrio_2015a} extended the analysis to a porous material with relatively small permeability where the inertial effects within the porous layer can be neglected. In this context, they highlighted the decoupled roles played by porosity and permeability. They also showed that porous media characterized by the same porous length scale in viscous units deliver the same effect on the outer turbulent flow. \citet{suga_matsumura_ashitaka_tominaga_kaneda_2010a} conducted an experimental investigation on the effects of wall permeability on laminar-turbulent transition in channel flows observing that the slip velocity over the permeable wall increases dramatically in the range of critical Reynolds numbers. In addition, consistently with the results of the linear stability analysis of \citet{tilton_cortelezzi_2006a, tilton_cortelezzi_2008a}, they found that the transition to turbulence appears at progressively lower Reynolds numbers as the value of the isotropic permeability is increased. They also showed that in a turbulent regime, the fluctuations of the normal-to-the-wall velocity component increase as the wall permeability and/or the Reynolds number increases. All the previously reported results were obtained considering porous isotropic media and only recently anisotropic permeable coatings have received some attention. In particular, a set of linear stability analysis reported in \citet{deepu_anand_basu_2015a} and in \citet{gomez-de-segura_sharma_garcia-mayoral_2017a} have shown that anisotropic porous layers may provide an effective mean for the passive control of turbulence transition being able to modulate the value of the critical Reynolds number for parallel wall bounded flows. Preliminary simulations of wall bounded turbulent flows have also shown a potential for drag reduction  \citep{gomez-de-segura_sharma_garcia-mayoral_2017a}. In particular, by means of stability analysis, the authors show that relaxing wall permeability triggers a Kelvin-Helmholtz instability which is responsible for the appearance of elongated spanwise vorticity structures. They also provide an upper bound for the achievable maximum drag reduction and, using preliminary DNS results, they show that anisotropic coatings may palliate the drag increase produced by isotropic ones. \citet{abderrahaman-elena_garcia-mayoral_2017a} conducted a detailed a priori analysis to assess the potential of these surfaces, and predicted a monotonic decrease in skin friction as the streamwise permeability increases. Very recently, \citet{kuwata_suga_2017a} used a DNS based on a Lattice Boltzmann MRT method to simulate a turbulent channel flow over a porous layer. The directional permeabilities are modulated by assembling arrays of hollows cubes along the Cartesian directions. In particular, they have considered four combinations obtained by complementing the permeability in the wall normal direction with added permeabilities along the streamwise and spanwise axes. Similarly to our findings, they also report strong effects on the structure of wall turbulence even at low permeability values. However, differently from the present study they did not consider the case in which wall normal permeability is reduced and, at the
same time, the cross plane one is enhanced.  We will show that this choice of permeability ratio, leads to a drag reducing condition.

In this work, we present  Direct Numerical Simulations (DNS) of a plane turbulent channel flow at a constant bulk Reynolds number  ($Re_b=2800$) bounded by two porous slabs with the same  porosity value ($60 \%$). A set of flow realisations is obtained by modifying the permeability of the porous media including isotropic cases and anisotropic ones with different entries in the permeability tensor. It will be shown that porous walls characterised by an anisotropic permeability allow for an effective passive manipulation of the near-wall turbulent flow. In particular, by tailoring the directionality of the permeability tensor one can either obtain a significant drag increase or decrease with decreased or enhanced local mixing. Either of the two outcomes can be beneficial depending on the technological application under consideration (e.g., skin friction drag reduction or separation control in an aerodynamic scenario). As also highlighted by \citet{kuwata_suga_2017a}, the modulation of the directional values of the permeability leads to completely different turbulence topologies in the near wall region where modulation or destruction of velocity streaks depends on the enhancement or the suppression of Kelvin Helmholtz spanwise rollers.

\section{Formulation} \label{sec:formulation}
\begin{figure}
  \centering
  \includegraphics[width=0.37\textwidth]{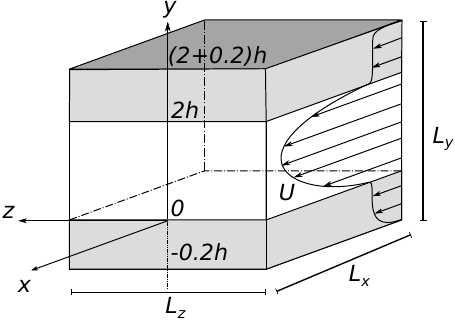} \hspace{1.0cm}
  \includegraphics[width=0.39\textwidth]{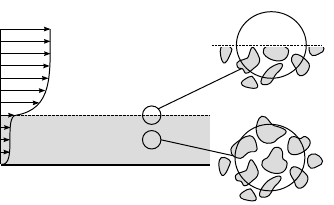}
  \caption{(a) Sketch of the channel geometry (not in scale), with interfaces between the fluid and the porous region located at $y=0$ and $2h$, and two solid rigid walls at $y=-h_p$ and $2h+h_p$. (b) Domain side view and sketch of the volume average process.}
  \label{fig:sketch}
\end{figure}
We consider a fully developed turbulent plane channel flow of an incompressible viscous fluid. The flow domain is delimited by two identical, flat, rigid, homogeneous, porous layers sealed by impermeable walls at their outer edges as shown in \figrefA{fig:sketch}. The lower and upper interfaces between the fluid and the porous material are located at $y = 0$ and $y=2h$, while the lower and upper impermeable walls are located at $y = -0.2h$ and $y= (2+0.2)h$ respectively.  As commonly done, we characterise the porous layer using the porosity $\varepsilon$ and the permeability $\overline{\overline{K}}$. The former is
a scalar quantity defined as the ratio of the volume of void over the total volume of a given material.
The latter is a symmetric second order, positive definite tensor that relates the velocity vector and the pressure gradient in a {\em Darcy}'s  porous medium. The entries in
the permeability tensor provide a direct measure of the ease with which a fluid flows through the medium along a certain direction. In the case of isotropic porous media, the off-diagonal terms are null and the diagonal ones assume the same value. 
This study will also consider anisotropic media characterised by a diagonal permeability tensor.
 In particular, the entries of the non-dimensional permeability tensor $\sigma_{ij}=\sqrt{K_{ij}}/h$ are taken to be null if $i \neq j$, $\sigma_{ij}=\sigma_{y}$ if $i=j=2$ and $\sigma_{ij}=\sigma_{xz}$ otherwise.
This choice is quite general since the permeability tensor is always diagonalizable along the principal directions with real, positive eigenvalues. We also introduce a characteristic Reynolds number as $Re_b = U_b h/\nu$, where $\nu=\mu/\rho$ is the fluid kinematic viscosity ($\mu$ and $\rho$ being the fluid dynamic viscosity and density, respectively), $h$ is half the layer to layer distance and  $U_b$ the bulk velocity. The latter is defined as the average of the mean velocity computed across the whole domain $U_b=\frac{1}{2(h+h_p)} \int_{-h_p}^{2h+h_p} \overline{u}(y) dy \simeq  \frac{1}{2h} \int_{0}^{2h} \overline{u}(y) dy$, with $h_p$ either set to zero when the wall is solid, or $0.2 h$ in the porous wall case \citep{breugem_boersma_uittenbogaard_2006a}. This choice facilitates the comparison with the canonical solid wall case since the contribution to the flow rate from inside the porous layer is almost negligible as the velocity magnitudes inside the layer are in the order of $10^{-4}$ of the bulk velocity. In the continuum limit, the flow of an incompressible viscous fluid through the whole domain that includes the porous layers, as sketched in \figref[a]{fig:sketch}, is governed by the Navier-Stokes equations. However, within the porous media, the imposition of the boundary conditions on the highly complex solid matrix and the related resolution requirements make this direct approach almost infeasible. To overcome these difficulties, \citet{whitaker_1969a, whitaker_1986a, whitaker_1996a} proposed to model only the large-scale behaviour of the flow in the porous medium by averaging the Navier-Stokes equations over a small sphere, of volume $V$ and radius $r$, as illustrated in \figref[b]{fig:sketch}. The introduction of the velocity field splitting  (\ie separation of large and small scales) and the application of the space averaging operator to the Navier Stokes equations leads to the so-called volume-averaged Navier-Stokes (VANS) equations. \cite{rosti_cortelezzi_quadrio_2015a}  have derived  a specific form of the VANS equations when the hypothesis of  isotropic porous medium, negligible fluid inertia and scale separation (\ie $\ell_s \sim \ell_f \ll r \ll L_p$,  $\ell_f$ and $\ell_s$ being the smallest scales of the fluid and the solid phase, and $L_p$ the characteristic thickness of the porous layer) are introduced. The extension of this formulation to  anisotropic porous media characterised by a permeability tensor of the form described above can be easily derived following the formal process introduced in \citet{rosti_cortelezzi_quadrio_2015a}. The resulting set of VANS equations tailored to this specific case reads as:
\begin{equation} 
\label{eq:VANS}
\dfrac{\partial \aver{\bm{u}}^s}{\partial t} = - \varepsilon \bm{\nabla} \aver{p}^f + \dfrac{1}{Re_b} \nabla^2 \aver{\bm{u}}^s - \dfrac{\varepsilon}{Re_b} \left( \dfrac{\aver{u}^s \bm{i}}{\sigma^2_{xz}} + \dfrac{\aver{v}^s \bm{j}}{\sigma^2_{y}} + \dfrac{\aver{w}^s \bm{k}}{\sigma^2_{xz}} \right), \; \bm{\nabla} \cdot \aver{\bm{u}}^s = 0,
\end{equation}
where $\bm{i}$, $\bm{j}$, and $\bm{k}$ are the unit vectors in the three coordinate directions. Equations (\ref{eq:VANS}) are obtained introducing two averaging operators over the volume of fluid $V_f$ contained within the elemental porous volume $V$: the superficial volume-average $\aver{\phi}^s = 1/V \int_{V_f} \phi dV_f$, and the intrinsic volume-average $\aver{\phi}^f = 1/V_f \int_{V_f} \phi dV_f$ ($\phi$ ia any fluid variable). The first average is linearly related to the second via the porosity value  \ie $\aver{\phi}^s = V_f/V \aver{\phi}^f = \varepsilon \aver{\phi}^f$. Because of the incompressible character of the flow, the superficial volume average is used for the velocity field, while the intrinsic volume average is preferred for the pressure being directly related to experimental values  \citep{quintard_whitaker_1994b, whitaker_1996a}. Equations (\ref{eq:VANS}) encompass both the solid and the fluid phase that are discriminated by the porosity value.
To close the formulation, we also need to introduce a proper treatment of the porous-fluid interface. For an uniform porous media 
we require pressure and velocity continuity at the interface, while the shear stress may show a jump at the interface \citep{ochoa-tapia_whitaker_1995a}. The magnitude of the jump discontinuity is controlled by a prescribed parameter $\tau$ that ultimately depends on the type of porous material considered and by the texture of the solid interface \citep{ochoa-tapia_whitaker_1998a}. The value of $\tau$ 
measures the transfer of stress at the porous/fluid interface \citep{minale_2014a,minale_2014b}: $\tau=0$ represents the situation in which the stress carried by the free flowing fluid is fully transferred to the fluid saturating the porous matrix, 
while non-zero values prescribe the portion of stress that can be hold by the solid matrix at the interface. \citet{rosti_cortelezzi_quadrio_2015a} have found that positive values of $\tau$ increase the slip velocity and decrease the Reynolds stresses at the interface, while zero and negatives values produce a drag increase that weakly depends on the value of $\tau$
itself. In this work, it will be assumed $\tau=0$, as a representative condition in which no drag reducing benefits are obtained (\ie $\tau=0$ produces a drag increase, when isotropic media are considered). Using the mentioned assumptions, the momentum-transfer conditions \citep{ochoa-tapia_whitaker_1995a} at the interfaces ($y=0$ and $y=2$) reduce to 
\begin{equation}
\label{eq:OTW_cond}
\bm{u} = \langle \bm{u} \rangle^s, \;\;\; p = \langle p \rangle^f, \;\;\; \left( \dfrac{\partial u}{\partial y} - \dfrac{1}{\varepsilon} \dfrac{\partial \langle u \rangle^s}{\partial y} \right) \bm{i} + \left( \dfrac{\partial w}{\partial y} - \dfrac{1}{\varepsilon} \dfrac{\partial \langle w \rangle^s}{\partial y} \right) \bm{k} = 0.
\end{equation}
It is worth noting that, imposing $\tau=0$ guarantees a general validity of the jump boundary condition \citep{minale_2014a,minale_2014b}. Moreover, the jump free condition on the shear stress that we have adopted corresponds to the one used by Minale. The validity of the latter has been experimentally verified by \citet{carotenuto_vananroye_vermant_minale_2015a}. Finally, the outermost portions of the  porous layers are bounded by  impermeable walls where no-slip and impermeability conditions are applied.

The governing equations are conveniently  made independent of pressure and reformulated in terms of wall-normal velocity and vorticity, $v$ and $\eta$ \citep{kim_moin_moser_1987a}. The derivation of the VANS in the poloidal-toroidal setting for the considered anisotropic medium can be easily obtained following \citet{rosti_cortelezzi_quadrio_2015a}.

\subsection{Numerical discretisation and simulations parameters}
The discrete counterpart of the governing equations \eqref{eq:VANS} and the interface conditions \eqref{eq:OTW_cond} are numerically tackled using a modified version of the code used for obtaining the results presented in \citet{rosti_cortelezzi_quadrio_2015a}. In what follows we shall give a short summary of the main numerical features of the code. Readers interested in further details can refer to \citet{rosti_cortelezzi_quadrio_2015a}. The governing equations are discretised in a box that is doubly periodic in the streamwise, $x$, and spanwise, $z$, directions. The spatial discretisation is spectral in  the $(x,z)$ planes using Fourier series while in the wall normal direction, fourth order compact finite-difference schemes are used. The time discretisation is based on a third order Runge-Kutta scheme for the nonlinear convective terms, which are computed in physical space using dealiasing in $(x,z)$ according to the $2/3$ rule, and on an implicit Crank-Nicolson scheme for the viscous and Darcy's terms. All the results that will be presented have been obtained on a single computational box which size on the $(x,z)$ plane is $[4 \pi h \times  2 \pi h]$. The periodic directions have been discretised using $256  \times 256$  Fourier modes before dealiasing. In the wall-normal direction, $150$ grid points are used in the fluid region while $75$ nodes are employed in each porous slab. In the $(x,z)$ planes the chosen discretisation in wall units (indicated by the superscript $^+$) leads to a spatial resolution of $\Delta x^+ \approx 8.8$ and $\Delta z^+ \approx 4.4$. The wall-normal resolution $\Delta y^+$ varies from $0.16$ by the interface to $4.1$ in the centerline region. Here the viscous wall units are defined using as friction velocity $u_{\tau}=\sqrt{\tau_{\mbox{\tiny total}}/\rho}$, being $\tau_{\mbox{\tiny total}}$ the total stress, \ie the sum of the viscous and the Reynolds stress evaluated at the fluid/porous interface  \citep{breugem_boersma_uittenbogaard_2006a}. Using $u_{\tau}$ as a velocity scale, we define the friction Reynolds number as $Re_\tau=u_\tau h/\nu$. Although the size of the computational domain and the spatial resolution are comparable to those employed by \citet{kim_moin_moser_1987a} at a similar bulk Reynolds number in a solid walls, plane turbulent channel flow, a further assessment on their adequacy has been checked a-posteriori. All the simulations have been initialised from the same field obtained from the DNS of a fully developed turbulent channel flow with solid walls. Initially, the simulations with porous layers have been advanced in time until the flow reached a statistical steady state. From this moment, the calculations are continued for an interval of $600 h/U_b$ time units, during which $120$ full flow fields are stored for further statistical analysis. To verify that the time interval was sufficiently long for accumulating statistically converged data, we have systematically  compared sequences of first and second order statistics, sampled during the integration time, to guarantee the achievement of a final satisfactory convergence.

\begin{table}
\centering
\setlength{\tabcolsep}{5pt}
\begin{tabular}{lrrrrrr}
Case												& $10^3~\sigma_{xz}$	& $10^3~\sigma_{y}$	& $\psi$	& $Re_\tau$	& $DR\%$	& $\overline{u}_i$ 	\\ 
\hline
I$\sigma\downarrow$					& $0.2500$					& $0.2500$					& $-$		& $181$	& $-3$	&	$0.0029$				\\ 
I$\sigma$										& $1.0000$					& $1.0000$					& $1$		& $182$	& $-6$	&	$0.0090$				\\ 
I$\sigma\uparrow$						& $4.0000$					& $4.0000$					& $-$		& $188$	& $-12$	&	$0.0384$				\\ 
\hline
A$\sigma_{xz}\downarrow\sigma_y\uparrow$																																		& $0.2500$					& $4.0000$					& $0.0625$	& $198$	& $-21$	&	$0.0034$				\\ 
A$\sigma_{xz}\uparrow\sigma_y\downarrow$																																		& $4.0000$					& $0.2500$					& $16$		& $177$	& $1$		&	$0.0336$			\\ 
A$\sigma_{xz}\uparrow\uparrow\sigma_y\downarrow\downarrow$																										& $8.0000$					& $0.1250$					& $64$		& $171$	& $8$	&	$0.0632$			\\ 
A$\sigma_{xz}\uparrow\uparrow\uparrow\sigma_y\downarrow\downarrow\downarrow$																			&	$16.0000$				& $0.0625$					& $256$	& $164$	& $18$	& $0.1149$			 
\end{tabular}
\caption{Summary of the DNSs performed at a fixed bulk Reynolds number equal to $Re=2800$ and with different porous layers, all with height $0.2h$ and porosity $\varepsilon=0.6$. The columns report the value of the wall-parallel $\sigma_{xz}$ and wall-normal $\sigma_{y}$ permeabilities, the anisotropy parameter $\psi=\sigma_{xz}/\sigma_y$, the friction Reynolds number $Re_\tau$, the drag reduction percentage $DR\%$ and the mean velocity at the interface $\overline{u}_i$, respectively.}
\label{tab:cases}
\end{table}
As already mentioned, all the simulations were carried out at the same bulk Reynolds number $Re_b=2800$.
In practice, we have considered flows sharing the same flow rate, thus facilitating the comparison with the baseline solid-walls case and also allowing to have a direct measurement of the total drag from the required mean pressure drop. 
The low value of the considered Reynolds number does not allow for an analysis for cases with
a clear scale separation, however it still provides a valuable insight of the effect of the
porous layers on the near wall flow.
As discussed by \citet{rosti_cortelezzi_quadrio_2015a}, when varying $Re$ (at least within a range of moderate to low values), the average flow variables near the interface still scale with viscous units provided that the permeability values are adapted to match the value of the porous Reynolds number $Re_K= \sqrt{\widetilde{K}} u_{\tau}/\nu$ ($\widetilde{K}$ is a measure of the magnitude of the permeability tensor, $\widetilde{K} = \max \left( {\sigma_{xz},\sigma_{y}} \right)$). Apart from the Reynolds number, the other parameters that are kept invariant throughout all the simulations are: the porosity of the slabs supposed to be homogeneous and the width of the porous layers $h_p$. In particular, we have chosen an intermediate value for the porosity (\ie $\varepsilon=0.6$), because its impact on the flow field is marginal as compared to the effects of permeability variations \citep{rosti_cortelezzi_quadrio_2015a}. We also have selected a porous layer thickness ($h_p=0.2 h$) which size is large enough to host a wide region where Darcy's regime is verified (\ie the transition region from the interface into the porous layer is small compared to the total thickness). \tabrefc{tab:cases} summarises the parameters defining all the simulations that have been carried out. Starting from the baseline case, denoted by I$\sigma\downarrow$, we have studied the effects of increasing permeability in the wall-normal (A$\sigma_{xz}\downarrow\sigma_y\uparrow$), wall-parallel (A$\sigma_{xz}\uparrow\sigma_y\downarrow$), and in all directions (I$\sigma\uparrow$). We have also studied the effect of anisotropy, by varying the value of the ratio $\psi=\sigma_{xz}/\sigma_y$. In particular, we have considered cases with permeability values assigned as $\sigma_{xz}=a\sigma$ and $\sigma_{y}=\sigma/a$, where $\sigma$ is the value of permeability of the isotropic case, and $a$ is a nondimensional constant constant set to the values $1/4$ (A$\sigma_{xz}\downarrow\sigma_y\uparrow$), $1$ (I$\sigma$), $4$ (A$\sigma_{xz}\uparrow\sigma_y\downarrow$), $8$ (A$\sigma_{xz}\uparrow\uparrow\sigma_y\downarrow\downarrow$), and $16$ (A$\sigma_{xz}\uparrow\uparrow\uparrow\sigma_y\downarrow\downarrow\downarrow$). 
 
It is worth mentioning that the highest values of the permeability have been bounded by a maximum limit to guarantee that inertia effect can be neglected both in the VANS equations and in the jump condition, \equref{eq:VANS} and \equref{eq:OTW_cond}, respectively. The validity of the former has been guaranteed by verifying that the ratio between the Forchheimer's coefficient (estimated via the Ergun's formula) and the Darcy's coefficient remains small. \citet{tilton_cortelezzi_2008a} showed that the Darcy's closure remains acceptable up to a permeability $\sigma=0.02$ which is above our maximum value. In particular, in all our simulations the ratio between the two terms remains below $10^{-3}$, thus rendering inertial effects negligible in the porous layers. The validity of the adopted interface condition is implicitly ensured by the choice of imposing the continuity of the stresses, \ie $\tau=0$. In this case, as discussed by \citet{tilton_cortelezzi_2008a}, inertial effects can be neglected in the interface shear matching condition without any limitation. As already mentioned, this choice turns out to be equivalent to the one proposed by \citep{minale_2014a,minale_2014b}, which has been experimentally verified by \citet{carotenuto_vananroye_vermant_minale_2015a}.

\subsection{Code validation}
\begin{figure}
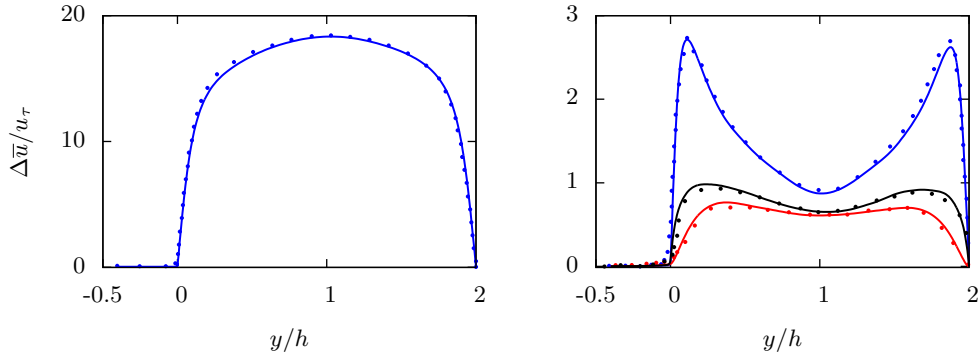

  \centering
  \input{fig02aL} \hspace{0.5cm}
  \input{fig02bL} \\ \vspace{0.8cm}
  \caption{(a) Mean streamwise velocity profile and (b) root mean square velocity profiles in wall units as a function of the wall-normal distance $y$. The figure shows the comparison of our numerical results (solid lines) with those reported by \citet{kuwata_suga_2017a} (symbols). The blue, red and black colours are used for streamwise, wall-normal and spanwise direction component, respectively.}
  \label{fig:val}
\end{figure}

To verify that the proposed formulation can be employed to tackle the simulation of turbulent flows over anisotropic porous media, we compare the predictions of the present method with the DNS-LMB results of \citet{kuwata_suga_2017a}. In particular, we consider the case of a plane channel that is bounded by an impermeable wall on one side and exposed to a porous medium on the opposite one. The porous medium is quasi impermeable in the $xz$ planes (i.e., we set the in-plane, non dimensional permeability $\sigma_{xz}=10^{-6}$ to avoid a division by zero) and has a wall-normal permeability equal to $\sigma_y=0.036$. The porosity of the medium is set to ($\varepsilon=0.56$). The bulk Reynolds number is set to $1775$  corresponding to a friction Reynolds number of $Re_\tau=111$. The left panel of \figrefAC{fig:val} shows the mean velocity profile, while the right one reports the distribution of the velocity fluctuations as a function of the wall-normal coordinate. An overall good agreement between the present solution, obtained by the volume averaged  \equref{eq:VANS}, and the reference one of  \citet{kuwata_suga_2017a} obtained using a direct wall-resolving numerical simulation is evident. It is worthwhile mentioning that this validation proves that the overall formulation that include the presented VANS (\equref{eq:VANS}) and the interface jump condition (\equref{eq:OTW_cond}) has the potential to reproduce to a high degree of fidelity the behaviour of turbulent flows over anisotropic media. It also shows that the numerical algorithm chosen for the discretization of the above equations is producing reliable results.

\begin{figure}
  \centering
  \includegraphics[width=0.33\textwidth]{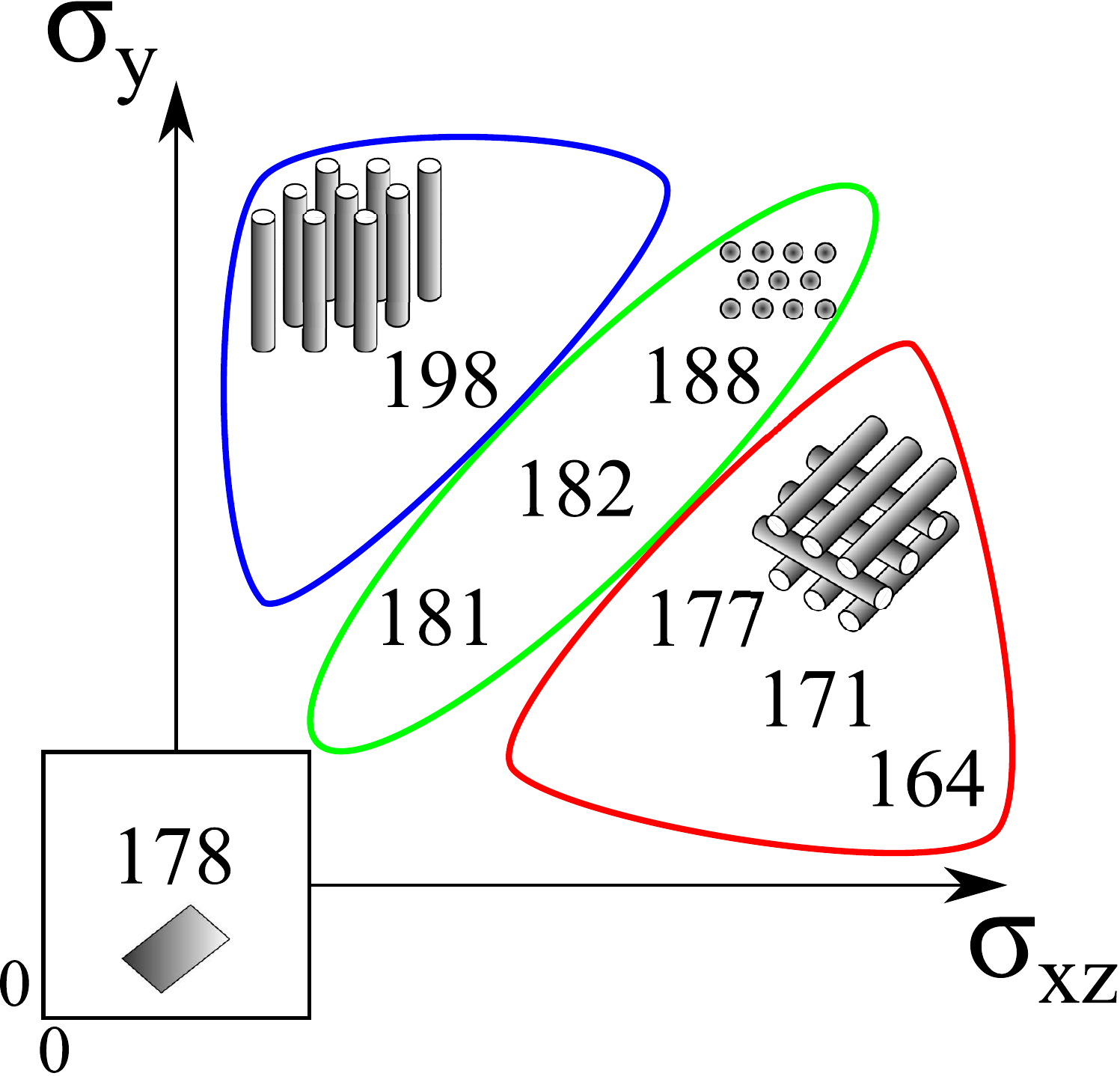} \hspace{1cm}
  \input{fig03bL}
  \vspace{0.5cm}
  \caption{(a) Symbolic representation of the friction Reynolds number $Re_\tau$ as a function of the permeability in the wall-normal $\sigma_y$ and wall-parallel direction $\sigma_{xz}$ (axis not to scale). The sketches in the left figure represent conceptual arrangements of porous material of various textures. (b) Profile of the drag reduction (black line) and the mean streamwise velocity $\overline{u}_i$ at the interface (grey line) as a function of $\psi$; blue, green, red, orange and magenta symbols are used for the cases A$\sigma_{xz}\downarrow\sigma_y\uparrow$, I$\sigma$, A$\sigma_{xz}\uparrow\sigma_y\downarrow$, A$\sigma_{xz}\uparrow\uparrow\sigma_y\downarrow\downarrow$, and A$\sigma_{xz}\uparrow\uparrow\uparrow\sigma_y\downarrow\downarrow\downarrow$, respectively.}
  \label{fig:global}
\end{figure}

\section{Results} \label{sec:result}
We start by considering the effect of the wall-normal and wall-parallel permeability on the mean friction coefficient. The friction Reynolds numbers $Re_\tau$, reported in \figref[a]{fig:global} and in \tabref{tab:cases} ($Re_\tau=178$ in the reference, impermeable case) grows to $181$ (I$\sigma\downarrow$) and $188$ (I$\sigma\uparrow$) in the isotropic cases when permeability is increased. The two anisotropic cases have instead a different behaviour: when the wall-normal permeability is greater than the wall-parallel one (A$\sigma_{xz}\downarrow\sigma_y\uparrow$), the friction Reynolds number further increases to $198$, while on the contrary when the wall-parallel component increases and the wall-normal reduces (A$\sigma_{xz}\uparrow\sigma_y\downarrow$), we recover a value close to the impermeable case (\ie $177$). Note that, the two anisotropic cases and I$\sigma\uparrow$ have the same permeability magnitude; from their comparison we can observe that lowering $\sigma_y$ or $\sigma_{xz}$ produces opposite effects on the $Re_\tau$ of almost the same amplitude. When the degree of anisotropy is increased (\ie higher $\psi$ values) a stronger drag reduction is observed (note that  \figref[a]{fig:global} also provide conceptual sketches of the considered porous materials). This effect is also visible in \figref[b]{fig:global}, that shows the percentage variation of the total drag $DR$ as a function of $\psi$. Here, the drag reduction variations are quantified by the ratio $DR=\left( G_0-G \right)/G_0$, where $G$ and $G_0$ are the mean pressure drops in the streamwise direction in the porous and solid wall cases, respectively. It is noted that the drag reduction increases monotonically with $\psi$ reaching a value of about $20\%$ within the considered range of permeability ratios (ie $\psi$ values that can be safely modeled with a Darcy's approach). The most striking feature of this graph is that the recorded drag reduction does not show any saturation level and that it could probably reach even higher values at higher $\psi$. This behaviour is remarkable when compared to other passive drag reduction techniques such as riblets that do present a drag reduction saturation with increased protrusion heights. \figrefC[b]{fig:global} also displays the mean slip velocity at the interface $\overline{u}_i$ as a function of $\psi$. Consistently with the behaviour of the variations of the drag reduction, the mean streamwise velocity on the interface monotonically increases within the range of considered $\psi$. Although a direct comparison with the results of \cite{kuwata_suga_2016a} is not possible (these authors considered the variations of all porous parameters, i.e.,  $\varepsilon$ and the three components of the permeability tensor), they also observe a similar behavior associated to the increase  of $\sigma_y$, \ie the friction Reynolds number $Re_\tau$ increases with $\sigma_y$.

\begin{figure}
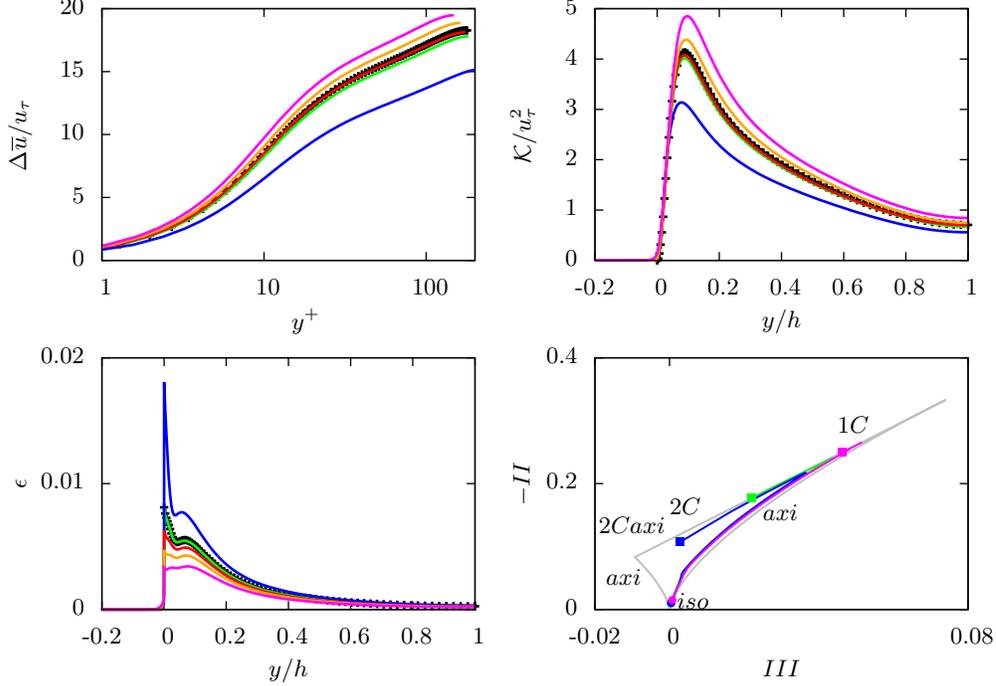

  \centering
  \input{fig04aL} \hspace{0.5cm}
  \input{fig04bL} \\ \vspace{0.8cm}
  \input{fig04cL} \hspace{0.5cm}
  \input{fig04dL}\\ \vspace{0.5cm}
  \caption{(a) Comparison of the streamwise mean velocity profile $u$ over rigid ($+$ symbols) and porous walls (solid lines) in wall units. 
Wall-normal profiles of (b) the turbulent kinetic energy $\mathcal{K}$ and of (c) the turbulent dissipation $\epsilon$. (d) The Lumley's triangle on the plane of the invariants $-II$ and $III$ of the Reynolds stress anisotropy tensor; the bounding grey line shows the region of admissible turbulence states, amid the extremal states of one-component turbulence (1C), two-components turbulence (2C), axisymmetric turbulence (axi), two-component axisymmetric turbulence (2C axi) and isotropic turbulence (iso). The bullets correspond to $y=h$ and the square to $y=0$. The colour scheme in all the graphs is the same as in \figrefA{fig:global}.}
  \label{fig:stat}
\end{figure}
\figrefC[a]{fig:stat} shows the profiles of the difference of the mean velocity and the interface velocity (\ie $\Delta \overline{u}=\overline{u}-\overline{u}_i$), normalised with the friction velocity, versus the logarithm of the distance from the interface measured in wall-units. The baseline, impermeable turbulent channel flow shows the presence of the three classical regions: the viscous sublayer for $y^{+}<5$ where the variation of $\overline{u}^{+}$ is approximately linear is followed by the log-law region, $y^{+}>30$. In between the two, a
buffer layer region (between $y^+ \simeq 5$ and $y^+\simeq30$) is also visible. \figrefC[a]{fig:stat}, shows that the velocity profiles from all simulations overlap in the viscous sublayer, depart in the buffer layer, and reach the equilibrium range with a logarithmic profile. The profiles reveal an upward shift in the log-law region (with respect to the impermeable walls case) in the drag-reducing cases (when $\psi>1$), and a downward shift when the drag increases.

The wall-normal distributions of the turbulent kinetic energy $\mathcal{K}=\overline{ u'_i u'_i} /2$  are shown in \figref[b]{fig:stat} together with the reference impermeable case. The profiles are strongly affected by the porous walls, especially close to the interface, where they take on non-zero values due to the relaxation of the no-slip condition on the permeable walls: a decrease in drag is accompanied by an increase of $\mathcal{K}$ at the wall. Increasing the anisotropy parameter $\psi$ also produces higher peaks with the maxima locations shifting towards the wall. Those increased values can be attributed to the amplification of the streamwise velocity fluctuations associated with an enhancement of the streaky structures above the walls, which will be further discussed. 

\figrefC[c]{fig:stat} displays the profiles of turbulent dissipation rate $\epsilon = \mu \overline{\partial_j u'_i \partial_j u'_i}$. The cases at $\psi>1$ are characterised by lower values of $\epsilon$ both at the wall and within a wide interior region. A different behaviour is observed in the other cases characterised by a drag increase ($\psi\le1$). Consistently with the decrease/increase of streamwise velocity fluctuations at the wall, decreased/increased viscous dissipations suggest a reduced/increased fragmentation of the flow structures (\ie suppression/enhancement of streaky structure). In all the cases, both the turbulent kinetic energy $\mathcal{K}$ and the dissipation $\epsilon$ drop to zero very quickly when entering the porous layer where the velocity fluctuations are mainly damped by the porous material itself.

\begin{figure}
  \centering
  \includegraphics[width=0.31\textwidth,angle=90]{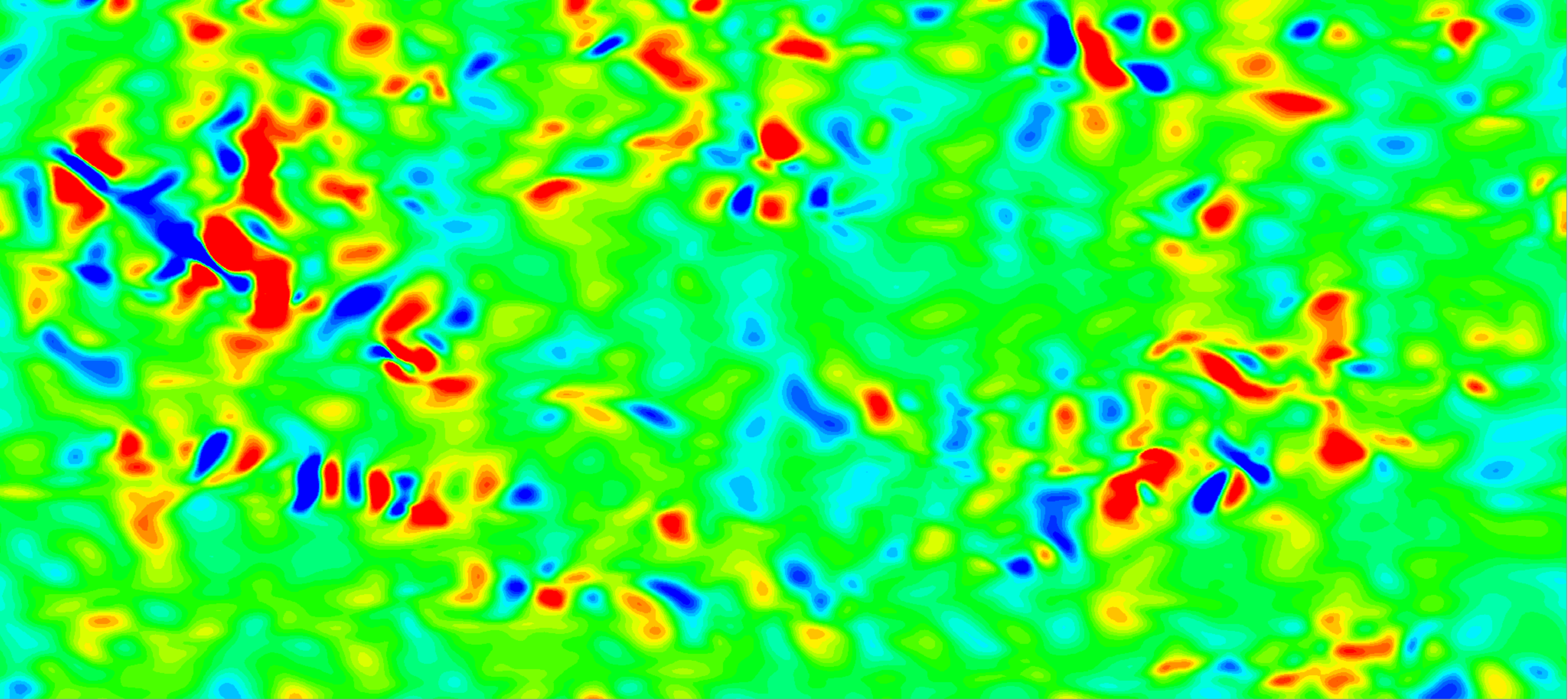}
  \includegraphics[width=0.31\textwidth,angle=90]{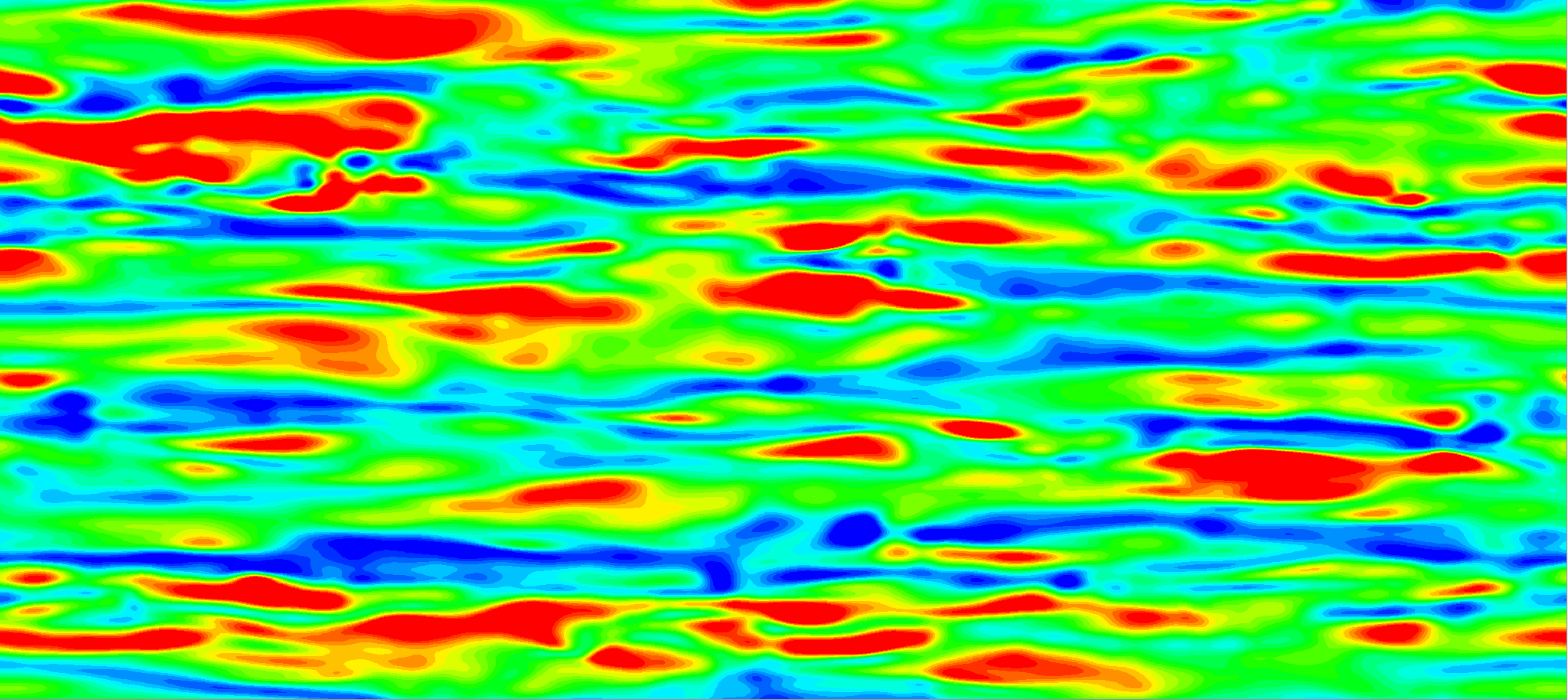}
  \includegraphics[width=0.31\textwidth,angle=90]{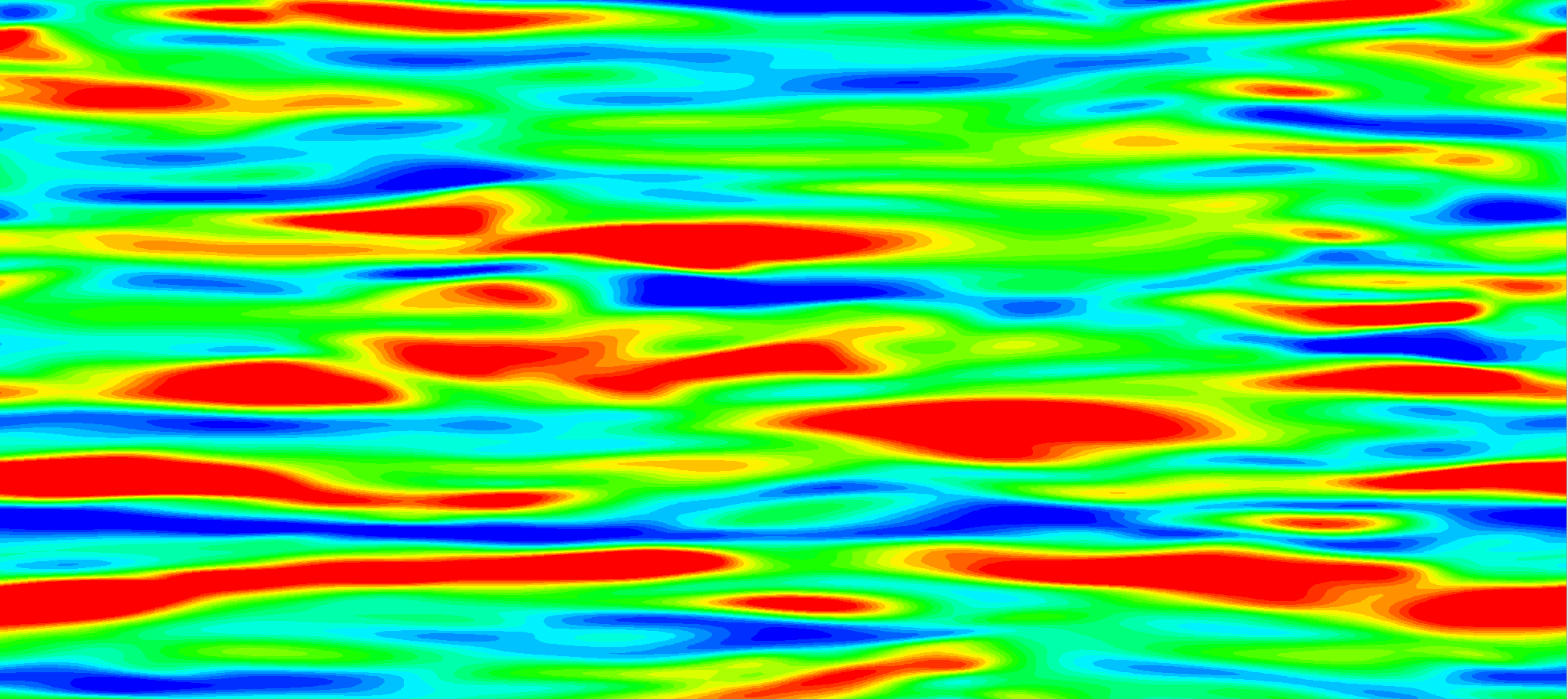} \hspace{0.5cm}
  \includegraphics[width=0.48\textwidth]{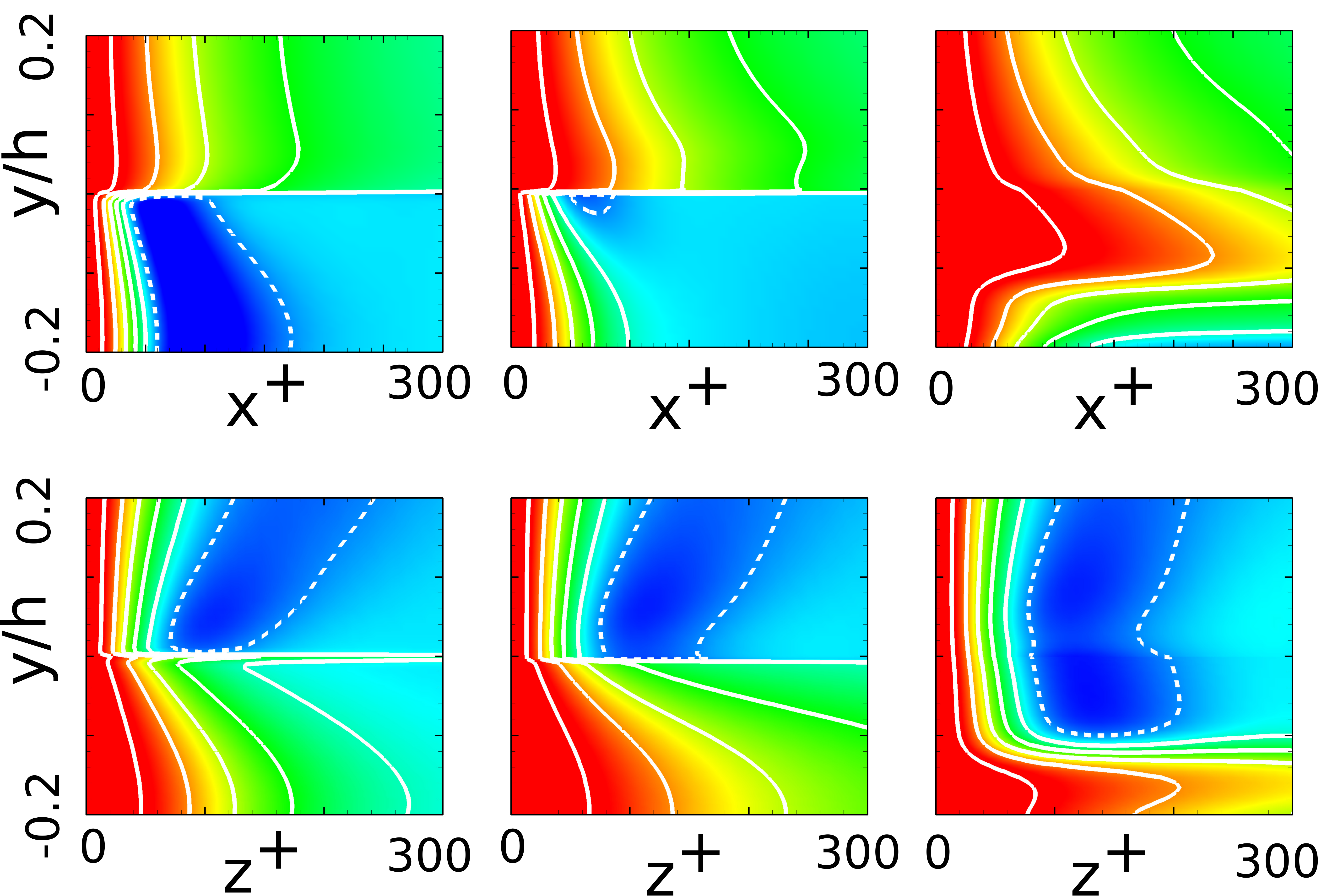}
  \caption{(a) Contours of instantaneous streamwise velocity fluctuation at the interface $y=0$ (flow going from bottom to top). The colors range from $-0.4\overline{u}_i$ (blue) to $0.4\overline{u}_i$ (red). (b-top) Streamwise velocity autocorrelation as a function of  $x^+$; (b-bottom) wall-normal velocity autocorrelation as a function of  $z^+$. The solid and dashed lines correspond to positive and negative values, ranging from $-0.1$ to $0.9$ with a step of $0.2$. The color scale ranges from $-0.2$ (blue) to $1.0$ (red). In each block, the figures refer to the drag-increasing A$\sigma_{xz}\downarrow\sigma_y\uparrow$, isotropic I$\sigma$, and drag-reducing A$\sigma_{xz}\uparrow\uparrow\uparrow\sigma_y\downarrow\downarrow\downarrow$ cases ($\psi$ increases from left to right).}
  \label{fig:struct}
\end{figure}
The variations in the near-wall flow topology are clearly observable in the left panel of \figrefA{fig:struct}, which displays instantaneous iso-contours of the streamwise component of the velocity at the interface for the cases of anisotropic ($\psi=0.0625$ and $\psi=256$) and isotropic porous layers. In all cases the high and low velocity regions are the footprints of the streaky pattern of the flow in the buffer layer. However, drag decreasing conditions show much wider and coherent structures, while the drag increasing case shows much less coherence with largely fragmented spotty contours. The break up of coherent close-to-the-wall structures induced by the presence of isotropic porous walls, or even by riblets at high protrusion heights has been observed in a number of other studies and often related to the insurgence of large spanwise rollers. These are caused by a Kelvin-Helmholtz instability induced by the presence of an inflectional point in the mean velocity profile \citep{jimenez_uhlmann_pinelli_kawahara_2001a, garcia-mayoral_jimenez_2011b} ultimately due to the relaxed condition on the vertical velocity component at the wall. At higher  $\psi$ values, the porous interface becomes almost impermeable and the appearance of spanwise rollers, which are the eigenfunctions of the most unstable modes, is now prohibited \citep{jimenez_uhlmann_pinelli_kawahara_2001a}.

The effect of the porous layer anisotropy on the flow coherence can be quantified by looking at the two-point velocity auto-correlation functions corresponding to the 
flow realisations of \figref[a]{fig:struct}. The top row in \figref[b]{fig:struct} shows the  distribution in the $x-y$ plane of the streamwise velocity component auto-correlation along the streamwise direction $x$.  
In the case $\psi=1$ (the isotropic case, shown in the central panel), the correlations appear to be very similar to the baseline, solid wall case with highly elongated streaky structures that alternate at a canonical spanwise distance  (\ie $\Delta z^+\simeq 100^+$) \citep{kim_moin_moser_1987a}. When the anisotropy parameter $\psi$ is increased, the correlation length increases monotonically and the velocity streaks penetrate more and more into the porous layer (see the rightmost panel) also revealing a less meandering shape (see \figrefA{fig:struct}). On the contrary, in the drag increasing scenario (\ie $\psi<1$) the $x$-wise correlations drops to zero more rapidly (see the leftmost panel) confirming a break-up of the streamwise coherence. The panels in the second row of \figref[right]{fig:struct}, showing the spanwise autocorrelation function of the normal velocity component, reveal that in the drag increasing scenario a long spanwise correlation length of the normal velocity establishes in the interface region. The appearance of this weak but coherent upwash and downwash motions may be related with an incipient emergence of the already mentioned spanwise rollers that would lift and splash the velocity streaks by the wall thus disrupting their coherence \citep{jimenez_uhlmann_pinelli_kawahara_2001a}.

In summary, when $\psi>1$ (\ie high wall-parallel and low wall-normal permeability), the flow is strongly correlated in the streamwise direction delivering a net drag decrease. On the other hand, when $\psi<1$ (\ie high wall-normal and low wall-parallel permeability), the flow tends to become more isotropic because of the fragmentation of the elongated streamwise streaks and the eventual  formation of new spanwise coherent structures. This type of phenomenon has been reported by other authors dealing with turbulent flows bounded by an isotropic permeable wall  \citep{jimenez_uhlmann_pinelli_kawahara_2001a, garcia-mayoral_jimenez_2011b, rosti_brandt_2017a}. \figrefC[d]{fig:stat} is a further evidence of the flow coherence modifications induced by the variations of the porous media permeability. 
In the diagram, the flow topology is clearly indicated by the shape of the Lumley's invariant maps computed for the isotropic layer and the most drag-reducing and most drag-increasing anisotropic cases. While the map for the  $\psi >1$ case (\ie the drag reducing condition)  shows a trend towards a one-component behaviour as the interface is approached, the $\psi \le 1$ cases show a modification towards a two-components turbulence behaviour.

\section{Conclusion} \label{sec:conclusion}
We have numerically investigated the effect of the porous media anisotropy on a plane turbulent channel flow bounded by porous layers. For a fixed, low Reynolds number and an intermediate porosity value, the flow inside the fluid region is tackled via a standard DNS, while a volume averaged approach is employed inside the porous layer. The two sets of equations are coupled imposing velocity and stress continuity at the interface. We showed that total drag over an anisotropic porous wall can be reduced or enhanced at least as much as $20\%$ by varying the ratio of the parallel to the wall and normal permeabilities $\psi$. A drag reducing behaviour is achieved whenever $\psi>1$, while a drag increase is recorded for $\psi\le 1$. Differently from other passive control strategies (\ie riblets), the drag reduction does not seem to saturate in the range of $\psi$ values that we have considered (in the case of riblets drag reduction saturates with the protrusion height). This behaviour suggests that porous material exhibiting high $\psi$ values, e.g. a grid of rods parallel to the streamwise and spanwise directions (\`a la {\em millefeuille}), may have a potential for even larger drag reductions. Such materials would reduce the wall-normal fluctuations while providing an enhanced slip velocity, thus leading to an overall decrease of the total interface stress. 
The reduced permeability also prevents the emergence of spanwise rollers induced by an inflectional instability of the mean velocity profile at the interface. On the other hand, porous materials with high wall-normal  and low wall-parallel permeabilities, \eg a carpet of wall-normal rods, are characterized by an increased turbulence isotropy in the near-to-the-interface region with a consequent disruption of the streamwise coherence probably due to the emergence of alternating spanwise correlated structures also leading to an increased drag.

The use of anisotropic porous coatings may open new ways to manipulate wall bounded turbulence. The main advantage of this approach is twofold: it is a passive technique that does not require the input of any external power, and it provides a self-regulating effect:  
the structure of the close-to-the-wall turbulence and in particular the drag reduction 
can be conserved if a change in the outer Reynolds number is compensated by a 
rescaling of the porous Reynolds number via a proper permeability scaling.

Although a practical design of the solid matrix that would deliver permeability ratios as those used here has not been envisaged yet, the present result complement the recent ones by \citet{kuwata_suga_2017a} who have modelled the anisotropy of the medium by stacking cubic open elements. The most important result of the present research, not covered by the mentioned work,  is a conceptual demonstration of the possibility of assembling a porous medium that can deliver large drag reductions in internal and external turbulent flows.

\section*{Acknowledgment}
This work was supported by the ERC-2013-CoG-616186 TRITOS, and by the VR 2014-5001. A.~P.~was supported by the UK-EPSRC grant EP/N01877X/1 {\em Quiet Aerofoils of the Next Generation}. The authors acknowledge the computer time provided by SNIC.

\bibliographystyle{jfm}
\bibliography{../../../../Articles/bibliography.bib}
\end{document}